\documentstyle [12pt,epsf] {article}

\begin{document}
\begin{center}
{\Large \bf Light Scalars in Cosmology\\}
\vspace{1.5in}
{\bf R. D. Peccei\\}
\vspace{0.1in}
{ \sl Department of Physics and Astronomy, UCLA, Los Angeles, CA 
90095-1547\\}
\vspace{1.5in}
\end{center}
\begin{abstract}
I discuss here some of the constraints imposed by quantum and gravitational
corrections on two hypothetical excitations, axions and quintessence,
which have important cosmological implications.  Although these corrections
can be kept under control, the resulting constraints are not too natural.
In particular, to keep the quintessence field light one must essentially
decouple it from ordinary matter.  Some possible suggestions of how to
avoid these troubles are briefly touched upon.
\end{abstract}

\newpage

\section*{Introductory Comments}

Everybody would agree that the structure of the Universe
follows from physical law.  Yet, often, in practice there is some disconnect 
in trying to relate cosmology to physics and {\it vice versa}.  In this talk
I want to raise some of these issues in the specific context of the role which
light scalar (or pseudoscalar) states play in cosmology.  However, I believe
that the lessons learned from this particular example are broader.

 I want to discuss specifically two types of light excitations which, if they
were to exist, would undoubtedly play a substantial cosmological role.  The
first of these particles is the axion \cite{RDP1}, whose origins have a very
good particle physics pedigree \cite{PQ}.  Axions are an excellent candidate
for the cold dark matter in the Universe if the scale of the $U(1)_{\rm PQ}$
symmetry breaking $f_a$ is large enough ($f_a \sim 10^{12}$ GeV).  The second
of these excitations is associated with quintessence \cite{Stein}, which
provides a possible explanation for the dark energy which appears to permeate the 
Universe.  In this case, the quintessence field $\phi$, provided it has negative
pressure $[\omega = p/\rho < 0]$, explains the apparent acceleration of the Universe and hence has a very good cosmological pedigree.  The relevant question which I want to pose here is how well does quintessence fare from a particle physics point of view.

Both axions and quintessence are associated with extremely light excitations.
The axion mass is inversely proportional to the $U(1)_{\rm PQ}$ breaking
parameter $f_a$: $m_a \simeq 6[10^6~{\rm GeV}/f_a]$ eV.  Hence if $f_a\sim
10^{12}$ GeV, so that axions help close the Universe, one is dealing with
axion masses in the micro eV range.  For quintessence the relevant dynamical
equation for the field $\phi$ ties the mass of the quintessence field to the present energy
density in the Universe: $\mu^2_\phi M_{\rm P}^2\sim (5\times 10^{-3}$ eV)$^4$, where $M_{\rm P}$ is the Planck mass.  Hence $\mu_\phi \sim 10^{-31}$ eV.  

For both axions and quintessence there are three general questions that needs
answering:

\begin{description}
\item{i)} How can one  physically keep these states so light?
\item{ii)} What is the role that gravitational interactions play in these
considerations?
\item{iii)} Are these states associated with other physical phenomena, besides
the ones which they were invented to solve?
\end{description}

As we shall see, the answer to the first query is that this is very hard
to do!  Indeed, when one considers the effects of gravity the results are 
generally catastrophic.  However, if one can keep these states light and
mute the effects of gravity, both axions and quintessence can lead to other
very interesting physical effects which can independently help confirm their existence.

\section*{Axion Musings}

For axions, the standard answer of why these states are light is that they
are pseudo Goldstone bosons.  The $U(1)_{\rm PQ}$ global symmetry is
spontaneously broken at a scale $f_a$, so that one normally would expect an
associated massless Nambu-Goldstone boson-- the axion.  However, $U(1)_{\rm PQ}$ is also an anomalous symmetry so that its symmetry current is not
divergenceless:
\begin{equation}
\partial_\mu J^\mu_{\rm PQ} = \frac{\alpha_s}{8\pi} G_a^{\mu\nu}
\tilde G_{a\mu\nu} + \kappa_{a\gamma\gamma}\frac{\alpha}{2\pi} 
F^{\mu\nu}\tilde F_{\mu\nu}~.
\end{equation}
As a result, the axion picks up a small mass which is of order
\begin{equation}
m_a\sim\frac{\Lambda^2_{\rm QCD}}{f_a}~,
\end{equation}
reflecting the dynamical scale at which the anomaly becomes effective\footnote{More precisely, $m_a\sim m_\pi f_\pi/f_a\sim m_q\Lambda^{3/2}_{\rm QCD}/f_a$.  Thus the
axion mass vanishes as the quark masses vanish, because another chiral
symmetry is restored.}

The effect of the anomaly and the concomitant mass generation for axions
results in an effective interactions of axions with gauge fields and an
associated axion effective potential \cite{RDP1}
\begin{eqnarray}
{\cal{L}}_{\rm eff} &=& \frac{a}{f_a}\left[\frac{\alpha_s}{8\pi} G_a^{\mu\nu}
\tilde G_{a\mu\nu} + \kappa_{a\gamma\gamma}\frac{\alpha}{2\pi} F^{\mu\nu}
\tilde F_{\mu\nu}\right] \\
V(a) &=& -\Lambda^4_{\rm QCD} \cos\frac{a}{f_a}~.
\end{eqnarray}
From the axion effective potential $V(a)$ one learns both that the axion picks
up a mass and that the vacuum expectation value  of the axion field vanishes,
corresponding to a vanishing effective vacuum angle:
\begin{equation}
\theta_{\rm eff} = \frac{\langle a\rangle}{f_a} = 0~.
\end{equation}

The above discussion, however, neglects gravitational effects.  One can
appreciate the possible troublesome role of gravity in this case by
considering the black hole ``no hair" theorem \cite{Banks}.  This theorem,
roughly speaking, states that black holes can only be characterized by a few
parameters (mass, spin, charge) but {\bf not} by global charges.  Because
black holes can absorb global charges, like $Q_{\rm PQ}$, and, at the same time, cannot
be characterized by these charges, it follows that gravitational
interactions must explicitly break the $U(1)_{\rm PQ}$ symmetry. 
 Schematically, this will lead to corrections to the Lagrangian of the theory by terms inversely proportional to the Planck mass \cite{Gbreak}. 
\begin{equation}
{\cal{L}}_{\rm eff} = {\cal{L}}_{\rm PQ} + \sum_M \frac{1}{M_{\rm P}^n}
O_n
\end{equation}
where $O_n$ are $U(1)_{\rm PQ}$-breaking operators of dimension $(n+4)$.

These additional gravitational corrections have significant effects,
altering drastically the naive axion properties, unless they are somehow
controlled  \cite{Gbreak}.  It is useful to illustrate what happens with a
simple example  \cite{RDP2}.  Consider for this purpose a modified axion
potential resulting from the presence of one such dimension  $(n+4)$ term:
\begin{equation}
V(a) = -\Lambda^4_{\rm QCD} \cos\frac{a}{f_a} - K\frac{f_a^{n+4}}{M_{\rm P}^n}
\cos\left(\frac{a}{f_a} + \delta\right)~.
\end{equation}
It is easy to see that the presence of the second term above both shifts the
axion mass and produces a non-zero $\theta_{\rm eff}$:
\begin{eqnarray}
m_a^2 & \simeq & \frac{\Lambda^4_{\rm QCD}}{f_a^2} + K\frac{f_a^{n+2}}{M_{\rm P}^n} \\
\theta_{\rm eff} &=& \frac{\langle a\rangle}{f_a} \simeq K\sin\delta
\frac{f_a^{n+4}}{M_{\rm P}^n\Lambda^4_{\rm QCD}}~.
\end{eqnarray}

Because $f_a$ is relatively near $M_{\rm P}$,  in general these effects are very large.  To preserve the Peccei-Quinn solution \cite{PQ} to the strong
CP problem one requires that the induced effective angle be very small,  $\theta_{\rm eff} < 10^{-10}$.  This happens either
if $\sin\delta=0$, so that CP is conserved in the $U(1)_{\rm PQ}$ breaking
interactions, or if
\begin{equation}
K\left[\frac{f_a^{n+4}}{M_{\rm P}^n\Lambda^4_{\rm QCD}}\right] < 10^{-10}~.
\end{equation}
Clearly the above obtains only if, somehow, the $U(1)_{PQ}$-violating interactions involve very high dimension operators
$(d>8)$, or if the coupling $K$ is extraordinarily small.  If either of the latter conditions holds, then the
axion mass shift is small.  However, if $\theta_{\rm eff}$ vanishes because
there is no CP violation in the  $U(1)_{PQ}$-breaking terms ( $\sin\delta=0$), then the LHS of Eq. (10) must be less than unity
to prevent a sizeable axion mass shift.

It is possible that nature conspires to mitigate the catastrophic effects which
gravitational interactions appear to have.  In particular, as has been argued
by Kallosh and collaborators \cite{Linde}, it may well be that the $U(1)_{\rm PQ}$
breaking interactions are naturally suppressed (in the langauge of the above
example $K \ll 1$). This perhaps is not that preposterous since, after all, black holes are extended objects and their effects on point-like interactions may well be exponentially suppressed.

Imagining that gravitational corrections are, somehow, under control,
opens up interesting possibilities.   In collaboration with Nussinov and Zhang \cite{NPZ}, recently I considered an intriguing example of this type of "controllable" correction. It involves a 
gravitationally induced $U(1)_{\rm PQ}$ violating contribution proportional
to the gluon field strengths.  Such a term, typified by the effective
Lagrangian
\begin{equation}
{\cal{L}}^{\rm PQ-viol}_{\rm eff} = \frac{f_a}{M_{\rm P}}\frac{\alpha_s}{4\pi}
G_a^{\mu\nu}G_{a\mu\nu}C\cos\left(\frac{a}{f_a} + \delta\right)~,
\end{equation}
gives rise to two correlated effects.  First, as before, such a term generates an
effective $\theta$-term. However, because $f_a\ll M_{\rm P}$ such a term does
not lead to too strong a constraint on $C\sin\delta$.  Indeed, since
\begin{equation}
\theta_{\rm eff}\simeq C\sin\delta\frac{f_a}{M_{\rm }}
\end{equation}
it follows that
\begin{equation}
C\sin\delta < 10^{-3}\left[\frac{10^{12}~{\rm GeV}}{f_a}\right]~.
\end{equation}

Secondly, the above interaction produces scalar modifications to gravity, due to
axion exchange.  The relevant term to consider here is the effective interaction
\begin{equation}
{\cal{L}}_{\rm eff} \simeq \frac{a}{M_{\rm P}} [C\sin\delta]
\frac{\alpha_s}{4\pi} G_a^{\mu\nu} G_{a\mu\nu}
\end{equation}
between axions and the gluon density.  Because the matrix element
of $G^2$ between nucleon states is essentially proportional to the nucleon
mass \cite{PSW}, the interaction (14) produces a Yukawa modification to the gravitational
potential coming from axion exchange.

The matrix element of $G^2$ between nucleon states can be inferred from the 
trace anomaly of the energy momentum tensor \cite{PSW}.   Focusing on the QCD
piece, one has
\begin{equation}
T_\mu^\mu = \frac{\beta(g_3)}{2g_3} G_a^{\mu\nu}G_{a\mu\nu} + m_q(1+\gamma_q)\bar qq~.
\end{equation}
Because
\begin{equation}
M_N = \langle N|T_\mu^\mu|N\rangle,
\end{equation}
and neglecting the small contribution coming from the quark mass term,\footnote{Interestingly these contributions give rise to composition dependent
forces. These effects, however, are not germane for the present considerations.} one finds that axion exchange lead to the following modification
to the gravitational potential:
\begin{equation}
V = V_{\rm grav}\left[1 + \alpha_a e^{-r/\lambda_a}\right]
\end{equation}
where
\begin{equation}
\alpha_a \simeq [C\sin\delta]^2 \leq 10^{-6}\left[\frac{10^{12}~{\rm GeV}}{f_a}\right]^2~. 
\end{equation} 
Note that for axions of cosmological significance the range of the force produced
from axion exchange is very short:
\begin{equation}
\lambda_a \simeq 2\left[\frac{f_a}{10^{12}~{\rm GeV}}\right]~cm. 
\end{equation}
Remarkably, these modifications are close to the present upper experimental
bounds of Hoskins {\it et al.} \cite{H} on modifications to gravity, which
for a range around a centimeter bound the strength of these additional forces to be
less than $10^{-4}$.

As a final comment, using Eq. (12) one sees that, in this example, there is
a natural hierarchy of interactions which the CP-odd axion has with
matter, with CP-even interactions strongly dominating over any CP-odd
interactions.  One has
\begin{equation}
{\cal{L}}^{\rm axion}_{\rm eff} = \frac{a}{f_a}
\left\{\frac{\alpha_s}{8\pi} G_a^{\mu\nu}\tilde G_{a\mu\nu} + \theta_{\rm eff}
\frac{\alpha_s}{4\pi} G_a^{\mu\nu} G_{a\mu\nu}\right\}~.
\end{equation}

\section*{Troubles for Quintessence?}

Many of the issues discussed above concerning axions apply {\it mutatis
mutandis} for quintessence.  Indeed, already some time ago Kolda and Lyth \cite{KL} remarked on the difficulty of keeping the mass of the quintessence field, $\mu_\phi$, light.  Also, Carroll \cite{Carroll} pointed out that, in general, one should expect gravitational strength interactions of quintessence with ordinary matter and discussed how these interactions can give rise to interesting microscopic and macroscopic phenomena.  Thus, my remarks here can
be relatively brief.

Keeping $\mu_\phi \sim 10^{-31}$ eV is a real problem.  In fact, without any
protective symmetry one gets enormous shifts.  Consider, for example,
the effective interactions of quintessence with electrons,
\begin{equation}
{\cal{L}}_{\rm eff} = \beta_e\phi\left(\frac{m_e}{M_{\rm P}}\right) \bar ee~.
\end{equation}
Even though this interaction is suppressed by the
Planck mass,
this term at one loop will generate a shift to the mass of the
quintessence field of order
\begin{equation}
\mu^2_\phi \to \mu^2_\phi + \beta_e^2 m_e^2.
\end{equation}
This is an enormous mass shift, unless one requires $\beta_e$ to be infinitesimal so as to keep $\mu_\phi$ at its tiny 
value.  Such disastrous effects are {\bf not} ameliorated by introducing
supersymmetry \cite{KL} since supersymmetry, if it exists, is broken
at the Fermi scale, not at a sub-eV scale!

One way out of this conundrum, which was suggested in the past \cite{Hill}, is to imagine that
the quintessence field is associated with some spontaneously broken global
symmetry which is only "very lightly" broken, either explicitly or through
anomalies.  However, as in the axion case, one again expects gravitationally
induced mass shifts which are not easily controlled.  So, even in this case,
one has again similar difficulties to those illuminated above.  Namely, if
one wants to keep $\mu_\phi$ very light, then necessarily the quintessence
field has only insignificant couplings to matter.

 Forgetting temporarely about these difficulties, Carroll \cite{Carroll} has constrained a variety of the possible couplings
that quintessence can have with gauge fields, through direct comparison with
experiment.  Writing the effective interactions for quintessence with these
fields as
\begin{equation}
{\cal{L}}_{\rm eff} = \frac{\phi}{M_{\rm P}}\left\{\beta_{G^2}G_a^{\mu\nu}
G_{a\mu\nu} + \beta_{F^2}F^{\mu\nu}F_{\mu\nu} + \beta_{F\tilde F}
F^{\mu\nu}F_{\mu\nu}\right\}~,
\end{equation}
Carroll\cite{Carroll} finds
\begin{equation}
\beta_{G^2} < 10^{-4}~;~~~\beta_{F^2} < 10^{-6}~;~~~
\beta_{F\tilde F} < 3\times 10^{-2}~.
\end{equation}

Although there are good prospects for improving these measurements in the
future,\footnote{For example, the limit on $\beta_{G^2}$ comes from looking for
composition-dependent forces and follows from equivalence principle tests at the
level of $\delta g/g \leq 10^{-12}$ \cite{Su}.  There are plans involving a
dedicated satellite experiment (STEP) to push these tests perhaps to the level of
$\delta g/g\sim 10^{-15}$.}  it is difficult to see how any of the $\beta_i$-parameters above can be anything but infinitesimal.  Indeed, our
simple example suggest that these couplings are of order $\beta_i\sim
10^{-40}$!  Also, as the discussion in the axion case made apparent, since
$\phi$ presumably is CP even one would expect $\beta_{F^2} \gg \beta_{F\tilde F}$.  Hence the good limit on $\beta_{F^2}$, obtained from limits on the variation of $\alpha$ with time coming from the Oklo natural reactor \cite{DD},
already suggests that one should have very small effects connected to the
coupling of quintessence to $F\tilde F$.  Hence, in reality, the very nice limit on $\beta_{F\tilde F}$
coming from the rotation of the plane of polarization of light coming from
distant sources \cite{Carroll} is probably not very meaningful.

\section*{Concluding Remarks}

Although light scalar fields play extremely interesting roles in cosmology,
unfortunately quantum and gravitational effects seem to argue against
their existence!  These arguments are more troubling for quintessence than axions, mostly because the quintessence field has to be so light.  Nevertheless, I believe one should not despair.  It is quite possible that
gravitational effects, although nominally suppressed by powers of the Planck mass, are actually much more strongly suppressed than that.  In particular, as mentioned earlier, the ``no hair" argument for the breaking of global symmetries involves extended objects. It is not hard to imagine that the finite size of black holes  could give  rise to a strong suppression factor \footnote{In the language of the simple example we discussed in the case of axions this corresponds to the statement that $K\ll 1$, so that all gravitational corrections are totally negligible.}

Furthermore, one should keep an open mind, concerning light states.
Even though now we do not understand how to stabilize these states to
radiative effects or gravitational shifts, it is possible that new ideas may
help.  For instance, recently Kogan {\it et al.} \cite{Kogan} argued for
the existence of light bulk states in theories with extra dimensions, with
the dynamics in the bulk helping to keep these states very light. There  is a corollary to this observation.
Because light scalar and pseudoscalar states through their coupling to matter
can cause interesting modifications to the standard lore, it is important to continue to
push the bounds one has on deviations from an $1/r^2$ gravitational force, and keep searching
for composition-dependent forces as well as for signs of time variability of
couplings.  If light scalars/pseudoscalars exist in nature, such phenomena
are possibly the most direct way to signal their presence in nature.

\section*{Acknowledgements}

This work was supported in part by the Department of Energy under
Contract No. FG03-91ER40662, Task C.

\end{document}